\documentclass[conference,10pt]{IEEEtran}

\usepackage{cite}

\usepackage{graphicx}
\usepackage{xcolor}
\usepackage{tikz}
\usetikzlibrary{patterns}
\usetikzlibrary{arrows,positioning,shapes.geometric,circuits.logic.US,spy}
\usepackage{pgfplots}
\usepackage{multirow}
\usepackage{upgreek}
\usepackage{amssymb}
\usepackage{amsmath}
\usepackage{cases}
\usepackage{url}
\usepackage{pifont}
\usepackage{bm}
\usepackage{amsthm}

\newtheorem{lemma}{Lemma}

\usepackage{tabularx}
\usepackage{booktabs}
\usepackage{filecontents}
\usepackage{subcaption}
\newcolumntype{Y}{>{\centering\arraybackslash}X}

\usepackage{algorithmic}

\usepackage{array}

\newcommand{\fixme}[2]{\ifx&#2&{\leavevmode\color{red}#1}\else{\leavevmode\color{red}FIXME\{}#1{\leavevmode\color{red}\}}\footnote{{\leavevmode\color{red}#2}}\PackageWarning{Fixme}{#1: #2}\fi}

\newcommand{\newstuff}[2]{\ifx&#2&{\leavevmode\color{blue}#1}\else{\leavevmode\color{blue}FIXME\{}#1{\leavevmode\color{blue}\}}\footnote{{\leavevmode\color{blue}#2}}\PackageWarning{Newstuff}{#1: #2}\fi}

\DeclareMathOperator*{\argmin}{arg\,min}
\DeclareMathOperator*{\sgn}{sgn}
\DeclareMathOperator{\PM}{PM}
\DeclareMathOperator{\modulo}{mod}
\DeclareMathOperator*{\arctanh}{arctanh}

\title{Generalized Fast Decoding of Polar Codes}

\author{Carlo~Condo, Valerio Bioglio, Ingmar~Land}

\begin{document}

\maketitle
\begin{abstract}
Research on polar codes has been constantly gaining attention over the last decade, by academia and industry alike, thanks to their capacity-achieving error-correction performance and low-complexity decoding algorithms. 
Recently, they have been selected as one of the coding schemes in the $5^{th}$ generation wireless standard (5G). 
Over the years various polar code decoding algorithms, like SC-list (SCL), have been proposed to improve the mediocre performance of the successive cancellation (SC) decoding algorithm for finite code lengths; however, like SC, they suffer from long decoding latency. 
Fast decoding of polar codes tries to overcome this problem by identifying particular subcodes in the polar code and decoding them with efficient decoders.
In this work, we introduce a generalized approach to fast decoding of polar codes to further reduce SC-based decoding latency. 
We propose three multi-node polar code subcodes whose identification patterns include most of the existing subcodes, extending them to SCL decoding, and allow to apply fast decoding to larger subsets of bits.
Without any error-correction performance degradation, the proposed technique shows up to $23.6\%$ and $29.2\%$ decoding latency gain with respect to fast SC and SCL decoding algorithms, respectively, and up to $63.6\%$ and $49.8\%$ if a performance loss is accepted, whose amount depends on code and decoding algorithm parameters, along with the desired speedup.
\end{abstract}

%

\IEEEpeerreviewmaketitle

\section{Introduction} \label{sec:intro}
Polar codes are a class of error-correcting codes proposed in \cite{arikan}. With infinite code length, under successive-cancellation decoding (SC), they can provably achieve capacity over binary memoryless channels. However, their error-correction performance degrades at finite code lengths, while SC yields long decoding latency, due to its inherent serial nature.

SC-list (SCL) decoding was proposed in \cite{tal_list}; it relies on a number of parallel SC decoders, and can substantially improve the error-correction performance of SC, especially when the polar code is concatenated with a cyclic redundancy check (CRC). This comes at the cost of additional complexity and latency. SC-flip decoding \cite{afisiadis} limits the increase in complexity by running a series of SC attempts, sacrificing decoding speed. Improved SC-based algorithms have led polar codes to be included in the $5^{th}$ generation wireless systems standard (5G) as one of the coding schemes for the enhanced mobile broadband communication scenario.

Different works over the years have attempted at reducing the decoding latency of SC-based algorithms at a limited complexity cost. The techniques described in \cite{alamdar,sarkis,hanif} rely on the recursive construction of polar codes to identify particular subcodes in their structure, and propose fast decoders for these subcodes that can be used in SC decoding. In \cite{hashemi_SSCL, hashemi_SSCL_TCASI,hashemi_FSSCL,hashemi_FSSCL_TSP,giardFlip}, the decoding of some of these subcodes has been extended to SCL and applied to SC-flip, reducing their latency and making them more practical to implement.

In this work, we introduce a generalized approach to fast decoding of polar codes to further reduce SC-based decoding latency. We propose three multi-node polar code subcodes whose identification patterns also envelop most of the existing subcodes \cite{alamdar,sarkis,hanif}. Moreover, we provide their extended decoding rules for both SC-based and SCL-based fast decoding: along with new subcodes, the codes identified in \cite{hanif} can thus be applied to SCL decoding. The impact of the proposed approach on SC and SCL decoding latency is evaluated, showing substantial speedup with respect to existing fast decoding algorithms \cite{sarkis, hashemi_SSCL_TCASI}. The error-correction performance loss brought by one of the new subcodes is then studied in terms of code, subcode and decoding algorithm parameters.

The remainder of this work is organized as follows. Section \ref{sec:prel} introduces polar codes, their decoding algorithms and existing fast decoding approaches. The proposed generalized fast decoding is detailed in Section \ref{sec:GFD}, while its speed and error-correction performance evaluation is carried out in Section \ref{sec:perf}. Conclusions are finally drawn in Section \ref{sec:conc}, along with future research directions.

\section{Preliminaries} \label{sec:prel}

\subsection{Polar Codes} \label{sec:prel:polar}

Polar codes are linear block codes of length $N=2^n$ and rate $R = K/N$.
They can be constructed using the transformation matrix $\mathbf{G}^{\otimes n}$ as
\begin{equation}
\mathbf{x} = \mathbf{u} \mathbf{G}^{\otimes n} \text{,} \label{eq:polarGen}
\end{equation}
that encodes vector $\mathbf{u} = \{u_0,u_1,\ldots,u_{N-1}\}$ into vector $\mathbf{x} = \{x_0,x_1,\ldots,x_{N-1}\}$.
The matrix $\mathbf{G}^{\otimes n}$ is obtained as the $n$-th Kronecker product of the polarizing kernel $\mathbf{G} = \left[\begin{smallmatrix} 1&0\\1&1 \end{smallmatrix}\right]$.

The polar encoding process identifies $K$ reliable bit-channels out of the $N$ available ones, and assigns the $K$ information bits to them. The remaining $N-K$ bit-channels in $\mathbf{u}$ are set to a known value, and represent the frozen set $\mathcal{F}$. 
To easily distinguish between frozen and information bits, a flag $s_i$ is assigned to each bit-channel, where
\begin{equation}
s_i = \begin{cases} 0 &\mbox{if } u_i \in \mathcal{F} \text{,} \\ 
1 & \mbox{otherwise }\text{.} \end{cases}
\end{equation}

\subsection{SC-Based Decoding} \label{sec:prel:SCDec}

\begin{figure}
  \centering
  \begin{tikzpicture}[scale=1.9, thick]
\newcommand\Triangle[1]{-- ++(0:2*#1) -- ++(120:2*#1) --cycle}
\newcommand\Square[1]{+(-#1,-#1) rectangle +(#1,#1)}

  \draw (0,0) circle [radius=.05];
  
  \draw (-.05,0) -- (.05,0);
  \draw (0,-.05) -- (0,.05);

  \draw (-1.05,-.55) \Triangle{.05};
  \draw (1,-.5) \Square{.05};

  \draw (-1.5,-1) circle [radius=.05];
  \draw (-.55,-1.05) \Triangle{.05};
  \draw (.5,-1) \Square{.05};
  \fill (1.5,-1) circle [radius=.05];

  \draw (-1.75,-1.5) circle [radius=.05];
  \draw (-1.25,-1.5) circle [radius=.05];
  \draw (-.75,-1.5) circle [radius=.05];
  \fill (-.25,-1.5) circle [radius=.05];
  \draw (.25,-1.5) circle [radius=.05];
  \fill (.75,-1.5) circle [radius=.05];
  \fill (1.25,-1.5) circle [radius=.05];
  \fill (1.75,-1.5) circle [radius=.05];

  \node at (-1.75,-1.7) {$\hat{u}_0$};
  \node at (-1.25,-1.7) {$\hat{u}_1$};
  \node at (-.75,-1.7) {$\hat{u}_2$};
  \node at (-.25,-1.7) {$\hat{u}_3$};
  \node at (.25,-1.7) {$\hat{u}_4$};
  \node at (.75,-1.7) {$\hat{u}_5$};
  \node at (1.25,-1.7) {$\hat{u}_6$};
  \node at (1.75,-1.7) {$\hat{u}_7$};

  \draw (0,-.05) -- (-1,-.45);
  \draw (0,-.05) -- (1,-.45);

  \draw (-1,-.55) -- (-1.5,-.95);
  \draw (-1,-.55) -- (-.5,-.95);
  \draw (1,-.55) -- (.5,-.95);
  \draw (1,-.55) -- (1.5,-.95);

  \draw (-1.5,-1.05) -- (-1.75,-1.45);
  \draw (-1.5,-1.05) -- (-1.25,-1.45);
  \draw (-.5,-1.05) -- (-.75,-1.45);
  \draw (-.5,-1.05) -- (-.25,-1.45);
  \draw (.5,-1.05) -- (.25,-1.45);
  \draw (.5,-1.05) -- (.75,-1.45);
  \draw (1.5,-1.05) -- (1.25,-1.45);
  \draw (1.5,-1.05) -- (1.75,-1.45);

  \draw [very thin,gray,dashed] (-2,0) -- (2,0);
  \draw [very thin,gray,dashed] (-2,-.5) -- (2,-.5);
  \draw [very thin,gray,dashed] (-2,-1) -- (2,-1);
  \draw [very thin,gray,dashed] (-2,-1.5) -- (2,-1.5);

  \node at (-2.25,0) {$t=3$};
  \node at (-2.25,-.5) {$t=2$};
  \node at (-2.25,-1) {$t=1$};
  \node at (-2.25,-1.5) {$t=0$};

  \draw [->] (-.12,-.05) -- (-1,-.4) node [above=-.1cm,midway,rotate=25] {$\bm{\alpha}$};
  \draw [->] (-.88,-.45) -- (0,-.1) node [below=-.1cm,midway,rotate=25] {$\bm{\beta}$};

  \draw [->] (-1.06,-.55) -- (-1.5,-.9) node [above=-.1cm,midway,rotate=40] {$\bm{\alpha}^{\ell}$};
  \draw [->] (-1.44,-.95) -- (-1.0,-0.6) node [below=-.1cm,midway,rotate=40] {$\bm{\beta}^{\ell}$};

  \draw [<-] (-.94,-.55) -- (-.5,-.9) node [above=-.1cm,midway,rotate=-40] {$\bm{\beta}^{\text{r}}$};
  \draw [<-] (-.56,-.95) -- (-0.975,-.625) node [below=-.1cm,midway,rotate=-40] {$\bm{\alpha}^{\text{r}}$};

\end{tikzpicture}
  \caption{SC-based decoding for a polar code with $N=8$, $R=1/2$.}
  \label{fig:SCDec}
\end{figure}
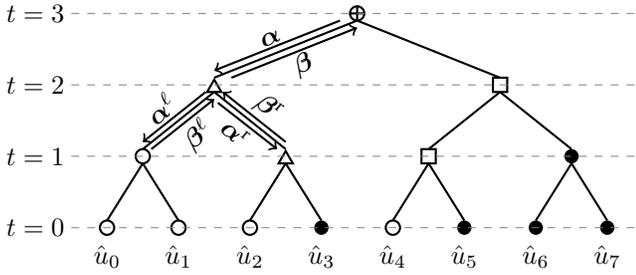

The SC decoding algorithm proposed in \cite{arikan} can be interpreted as a binary tree search, as portrayed in Fig.~\ref{fig:SCDec}. Every node at stage $t$ receives soft information in the form of logarithmic likelihood ratios (LLRs) from its parent node ($2^{t+1}$-element $\bm{\alpha}$ vector), and returns the hard decision vector $\bm{\beta}$. The tree is explored depth-first, with priority being given to the left branches.

The LLR vector $\bm{\alpha}^\ell$ sent to the left child node is computed through the $f_t$ function as
\begin{equation}
\label{eq:Ffunc1}
\alpha^{\ell}_i  =f_t(\alpha)= 2\arctanh \left(\tanh\left(\frac{\alpha_i}{2}\right)\tanh\left(\frac{\alpha_{i+2^{t}}}{2}\right)\right) \text{,} 
\end{equation}
where (\ref{eq:Ffunc1}) identifies the $f_t$ function. The LLR vector $\bm{\alpha}^\text{r}$ directed to the right child node is instead calculated through the $g_t$ function:
\begin{equation}
\label{eq:Gfunc1} 
\alpha^{\text{r}}_i = g_t(\alpha)= \alpha_{i+2^{t}} + \left(1-2 \beta^\ell_i \right) \alpha_i \text{.}
\end{equation}
Partial sums $\bm{\beta}$ are computed as
\begin{equation}
\beta_i =
  \begin{cases}
    \beta^\ell_i\oplus \beta^\text{r}_i \text{,} & \text{if} \quad i < 2^{t} \text{,}\\
    \beta^\text{r}_{i-2^{t}} \text{,} & \text{otherwise} \text{,}
  \end{cases}
  \label{eq:beta}
\end{equation}
where $\oplus$ is the bitwise XOR operation. At leaf nodes, $\beta_i$ is set to the estimated bit $\hat{u}_i$:
\begin{equation}
\hat{u}_i =
  \begin{cases}
    0 \text{,} & \mbox{if } s_i=0 \mbox{ or } \alpha_{i}\geq 0\\
    1 \text{,} & \mbox{if } s_i=1 \mbox{ and } \alpha_{i}< 0
  \end{cases} \label{eq:leafSC}
\end{equation}
The SC decoding process, and in particular the exploration of the tree according to its schedule, can be viewed as a sequence of $f_t$ and $g_t$ operations. For example, the exploration of the tree represented in Fig. \ref{fig:SCDec} can be expressed as $\{f_2,f_1,f_0,g_0,g_1,f_0,g_0,g_2,f_1,f_0,g_0,g_1,f_0,g_0\}$.

SCL decoding \cite{tal_list} maintains $L$ concurrent decoding candidates. At leaf nodes, $\hat{u}_i$ is estimated as both 0 and 1 when not a frozen bit, doubling the number of candidates. To limit the exponential increase in complexity, a path metric (PM) is assigned to each candidate \cite{balatsoukas_SCL_HW}:
\begin{align}
\PM_{{i}} = \begin{cases}
    \PM_{{i-1}} + |\alpha_{i}| \text{,} & \text{if } \hat{u}_{i} \neq \text{HD}(\alpha_{i})\text{,}\\
    \PM_{{i-1}} \text{,} & \text{otherwise,}
  \end{cases} \label{eq7}
\end{align}
where $\PM$ is initialized as 0 and
\begin{align}
\text{HD}(\alpha_{i}) = \begin{cases}
    0 & \text{if } \alpha_{i}\ge 0\text{,}\\
    1 & \text{otherwise.}
  \end{cases}
\end{align}
The $L$ candidates with the lowest PM are allowed to survive. SCL error-correction performance ca be further improved by concatenating the polar code with a CRC, that help in the selection of the final candidate.


\subsection{Fast SC-Based Decoding} \label{sec:prel:FPDec}

To increase the speed of SC-based decoding, in \cite{alamdar,sarkis,hanif}, particular sequences of frozen and information bits have been identified, and efficient fast decoders have been proposed. The decoding of the subcodes identified by these patterns, called special nodes, avoids the complete exploration of the SC-tree, allowing substantial speed increment.

Fast simplified SC decoding (Fast-SSC, \cite{sarkis}) considers four special nodes, whose structures are summarized as follows:
\begin{itemize}
\item \emph{Rate-0 Node}: all bits are frozen, $\mathbf{s} = \{0,0,\ldots,0\}$.
\item \emph{Rate-1 Node}: all bits are information bits, $\mathbf{s} = \{1,1,\ldots,1\}$.
\item \emph{Repetition (Rep) Node}: all bits are frozen except the last one, $\mathbf{s} = \{0,\ldots,0,0,1\}$.
\item \emph{Single parity-check (SPC) Node}: all bits are information bits except the first, $\mathbf{s} = \{0,1,1,\ldots,1\}$.
\end{itemize}
Additional special nodes and their efficient SC decoders have been observed in \cite{hanif}:
\begin{itemize}
\item \emph{Type-I Node}: all bits are frozen bits except the last two, $\mathbf{s} = \{0,\ldots,0,1,1\}$.
\item \emph{Type-II Node}: all frozen bits except the last three, $\mathbf{s} = \{0,\ldots,0,1,1,1\}$.
\item \emph{Type-III Node}: all information bits except the first two, $\mathbf{s} = \{0,0,1,\ldots,1\}$.
\item \emph{Type-IV Node}: all information bits except the first three, $\mathbf{s} = \{0,0,0,1,\ldots,1\}$.
\item \emph{Type-V Node}: all frozen bits except the last three and the fifth-to-last, $\mathbf{s} = \{0,\ldots,0,1,0,1,1,1\}$.
\end{itemize}

\section{Generalized Fast Decoding} \label{sec:GFD}

The nodes described in Section \ref{sec:prel:FPDec} identify patterns in the frozen and information bits. 
While in \cite{sarkis}, along with Rate-0, Rate-1, Rep and SPC nodes, some node mergers have been identified, literature and decoding methods have mostly focused on single node types. 
However, the identification of multi-node patterns leads to a generalized approach to fast decoding of polar codes. 
In this section, we describe three multi-node frozen and information bit patterns that can be exploited to increase the decoding speed at low complexity. 
They envelop Rep and SPC nodes, together with Type-I to V nodes, and extend their properties to a wider set of patterns. Moreover, general identification and decoding rules for both SC and SCL fast decoding are provided.

\subsection{Generalized Repetition Node} \label{subsec:GREP}

\begin{figure}
  \centering
  \begin{tikzpicture}[scale=1.9, thick]
\newcommand\Triangle[1]{-- ++(0:2*#1) -- ++(120:2*#1) --cycle}
\newcommand\Square[1]{+(-#1,-#1) rectangle +(#1,#1)}


  \node at (0,.25) {G-Rep};

  \fill (0,0) +(-0.05,-0.05) rectangle +(0.05,0.05);
  

  \draw (-1,-.5) circle [radius=.05];
  \draw (1,-.5) circle [radius=.05];
  
  \node at (-1,-.75) {Rate-0};  

  \draw (.5,-1) circle [radius=.05];
  \fill[gray] (1.5,-1) circle [radius=.05];

  
  \node at (.5,-1.25) {Rate-0}; 
  \node at (1.5,-1.25) {Rate-C}; 



  \draw (-1.75,.45) -- (0,.05);

  \draw (0,-.05) -- (-1,-.45);
  \draw (0,-.05) -- (1,-.45);

  \draw (1,-.55) -- (.5,-.95);
  \draw (1,-.55) -- (1.5,-.95);


  \draw [very thin,gray,dashed] (-2,.5) -- (2,.5);
  \draw [very thin,gray,dashed] (-2,0) -- (2,0);
  \draw [very thin,gray,dashed] (-2,-.5) -- (2,-.5);
  \draw [very thin,gray,dashed] (-2,-1) -- (2,-1);

  \node at (-2.25,.5) {$t+1$};
  \node at (-2.25,0) {$t$};
  \node at (-2.25,-.5) {$\vdots$};
  \node at (-2.25,-1) {$p$};
  \draw [->] (-1.55,.45) -- (0,.1) node [above=-.1cm,midway,rotate=-12] {$\bm{\alpha}$};
  \draw [->]  (-.12,0.02) -- (-1.75,.4) node [below=-.1cm,midway,rotate=-12] {$\bm{\beta}$};
%
  \draw [->] (1.06,-.55) -- (1.5,-.9) node [above=-.1cm,midway,rotate=-40] {$\bm{\beta}^{\text{r}}$};
  \draw [->] (1.44,-.95) -- (1.0,-0.6) node [below=-.1cm,midway,rotate=-40] {$\bm{\alpha}^{\text{r}}$};
  
%
%

\end{tikzpicture}
  \caption{Generalized Repetition Node}
  \label{fig:Grep}
\end{figure}
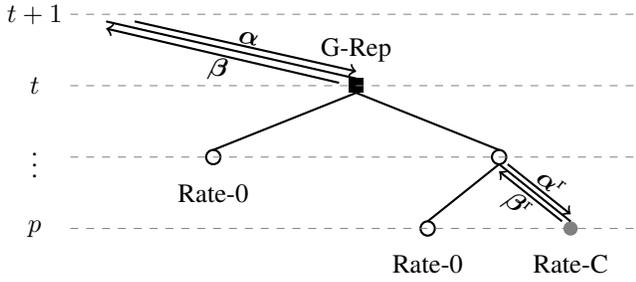

Repetition nodes are so named due to the pattern identified in the calculation of $\bm{\beta}$ (\ref{eq:beta}). 
In fact, vector $\bm{\beta}$ of a Rep node can be computed by performing a hard decision on the sum of the LLRs present in vector $\bm{\alpha}$ and replicating this result. 
However, this repetition pattern can be applied to a more general class of nodes. 
We call generalized Rep node (G-Rep) any node at stage $t$ for which all its descendants are Rate-0 nodes, except the rightmost one at a certain stage $p<t$, that is a generic node of rate C (Rate-C). 
The structure of a G-Rep node is depicted in Fig.~\ref{fig:Grep}. 
A G-Rep node can be decoded under SC using only the partial sum vector $\bm{\beta}^\text{r}$ of its Rate-C descendant and repeating it $2^{t-p}$ times. 

\begin{lemma}
The partial sum vector of a G-Rep node is given by $\bm{\beta} = \{ \bm{\beta}^\text{r},\dots,\bm{\beta}^\text{r} \}$, where $\bm{\beta}^\text{r}$ is calculated on the basis of the LLRs vector $\bm{\alpha}^\text{r}$ defined as $\alpha^\text{r}_i = \sum_{j = 0}^{2^p-1} \alpha_{i + j 2^{t-p}}$. 
\begin{proof}
If we call $P$ the list of operations needed to decode the Rate-C node, the list of operations needed to decode the G-Rep node is given by $\{ g_t,g_{t-1},\dots,g_{p+1},P \}$. 
By induction, the LLR vector $\bm{\alpha}^\text{r}$ of the Rate-C node is calculated as $\alpha^\text{r}_i = \sum_{j = 0}^{2^p-1} \alpha_{i + j 2^{t-p}}$ since all the left nodes are Rate-0 and hence $\beta^\ell_i=0$ in \eqref{eq:Gfunc1}. 
This vector is then used to decode the node performing the operations in $P$, that return the vector $\bm{\beta}^\text{r}$. 
The partial sum vector $\bm{\beta}$ at stage $p$ is then calculated recursively using \eqref{eq:beta} with $\beta^\ell_i=0$ by construction, obtaining that $\bm{\beta} = \{ \bm{\beta}^\text{r},\dots,\bm{\beta}^\text{r} \}$.
\end{proof}
\end{lemma}

According to the lemma, the output of a G-Rep node can be calculated as follows. 
First, the LLR vector $\bm{\alpha}^\text{r}$ of its Rate-C node is calculated as
\begin{equation}
\alpha^\text{r}_i = \sum_{j = 0}^{2^p-1} \alpha_{i + j 2^{t-p}}.
\end{equation}
Then, the Rate-C node is decoded under SC, obtaining the partial sum vector $\bm{\beta}^\text{r}$. 
In the case that the Rate-C node is a special node, partial sums can be computed through fast decoding techniques. 
Finally, the partial sum vector $\bm{\beta}$ of the G-Rep node is given by
\begin{equation}
\bm{\beta} = \underbrace{ \{ \bm{\beta}^\text{r},\dots,\bm{\beta}^\text{r} \} }_{2^{t-p}}
\end{equation}

Several special nodes identified in the past are particular cases of the G-Rep node class: aside from the straightforward Rep node, also Type-I, Type-II and Type-V nodes fit into this category, as long as all information bits are found in the rightmost child node.
In \cite{Ercan_SIPS17}, the G0R nodes proposed to merge operations in hardware implementations of fast-SSC decoders are G-Rep nodes for $t=p$, $t=p+1$ and $t=p+2$.

\subsection{Generalized Parity-Check Node} \label{subsec:GPC}

Frozen bits drive the decoding process thanks to their predetermined value: since they are assigned to low-reliability channels, bit estimations likely to be incorrect can be avoided and LLRs representing wrong values can be influenced positively. 
From an algebraic point of view, each frozen bit adds a constraint on the possible value of the codeword. 
Given the recursive nature of polar codes, the same concept applies to the decoding of constituent codes, or nodes in the SC tree. 
The constraint imposed by the frozen bit in SPC nodes is that of even parity in the codeword \cite{sarkis}: it can be exploited through Wagner decoding, i.e computing the parity of all the node bits and, if not fulfilled, flipping the one associated to the least reliable LLR. 
The two frozen bits in the leftmost positions of a Type-III node impose even parity constraints on the codeword, namely even and odd bit indices are treated as separate SPC nodes.  
Type-IV nodes rely on the same concept: the three frozen bits impose even parity constraints on bit indices modulo 4. 
However, since the fourth bit is an information bit, a suboptimal artifice is developed so that a parity constraint is imposed on the remaining bits, and four separate SPC nodes can be identified and decoded with Wagner decoding.
%
\begin{figure}
  \centering
  \begin{tikzpicture}[scale=1.9, thick]
\newcommand\Triangle[1]{-- ++(0:2*#1) -- ++(120:2*#1) --cycle}
\newcommand\Square[1]{+(-#1,-#1) rectangle +(#1,#1)}

  \node at (0,.25) {G-PC};

  \fill (-.05,-.05) -- ++(0:0.1) -- ++(120:0.1) --cycle;
  

  \draw (-1,-.5) circle [radius=.05];
  \fill (1,-.5) circle [radius=.05];
  
  \node at (1,-.75) {Rate-1};  

  \draw (-1.5,-1) circle [radius=.05];
  \fill (-.5,-1) circle [radius=.05];

  
  \node at (-1.5,-1.25) {Rate-0}; 
  \node at (-.5,-1.25) {Rate-1};   



  \draw (-1.75,.45) -- (0,.05);

  \draw (0,-.05) -- (-1,-.45);
  \draw (0,-.05) -- (1,-.45);

  \draw (-1,-.55) -- (-1.5,-.95);
  \draw (-1,-.55) -- (-.5,-.95);


  \draw [very thin,gray,dashed] (-2,.5) -- (2,.5);
  \draw [very thin,gray,dashed] (-2,0) -- (2,0);
  \draw [very thin,gray,dashed] (-2,-.5) -- (2,-.5);
  \draw [very thin,gray,dashed] (-2,-1) -- (2,-1);

  \node at (-2.25,.5) {$t+1$};
  \node at (-2.25,0) {$t$};
  \node at (-2.25,-.5) {$\vdots$};
  \node at (-2.25,-1) {$p$};

  \draw [->] (-1.55,.45) -- (0,.1) node [above=-.1cm,midway,rotate=-12] {$\bm{\alpha}$};
  \draw [->]  (-.12,0.02) -- (-1.75,.4) node [below=-.1cm,midway,rotate=-12] {$\bm{\beta}$};

%
  \draw [->] (-1.06,-.55) -- (-1.5,-.9) node [above=-.1cm,midway,rotate=40] {$\bm{\alpha}^{\ell}$};
  \draw [->] (-1.44,-.95) -- (-1.0,-0.6) node [below=-.1cm,midway,rotate=40] {$\bm{\beta}^{\ell}$};
%
%

\end{tikzpicture}
  \caption{Generalized Parity-Check Node}
  \label{fig:GPC}
\end{figure}
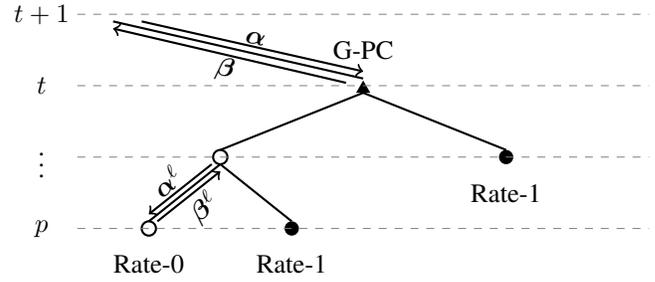

This even parity constraint can be generalized to a wider category of frozen bit patterns.
We call generalized parity-check node (G-PC) a node at stage $t$ having all Rate-1 descendants, except the leftmost one at a certain stage $p<t$ that is Rate-0. 
This structure, depicted in Fig.~\ref{fig:GPC}, imposes $N_p$ parallel single parity checks as follows. 
\begin{lemma}
A G-PC node at stage $t$ with the Rate-0 note at stage $p$ contains $N_p$ parallel SPC constraints such that $\sum_{i=0}^{2^{t-p}} \alpha_{i N_p+j} = 0$ for all $j = 0,\dots,N_p-1$.
\begin{proof}
A G-PC node identifies a code of rate $R = 1 - N_p/N_t$, for which $N_p$ bits out of $N_t$ are redundancy.  
So, if there exist $N_p$ independent parity check constraints, we have that those ones are the only constraints that should be used in the decoding, since all other constraints are linear combinations of these independent constraints.
The generator matrix of the code identified by the G-PC node is $M = (G^{\otimes t-p})_{0} \otimes G^{\otimes p}$, where $(G^{\otimes n})_{0}$ represents the matrix obtained by the $n$-Kronecker power of $G$ excluding the first row. 
All the rows of $(G^{\otimes t-p})_{0}$ have even weight by construction. 
This imposes an even parity check on the codewords of the code defined by $M$. 
More in detail, given $j\in\{0, ... , N_p-1\}$, the XOR of bits with index $i\modulo N_p=j$, $0\le i < N_t$ is equal to zero.
These parity check constraints are clearly independent, since they have no bits in common.
\end{proof}
\end{lemma}
The lemma suggests to decode a G-PC node with $N_p$ parallel Wagner decoders.
The LLRs vector $\bm{\alpha}$ of the G-PC node is thus divided into $N_p$ parts $\bm{\alpha^0},\dots,\bm{\alpha^{N_p-1}}$ such that
\begin{equation}
\alpha_i^j = \alpha_{iN_p + j}.
\end{equation}
Every LLRs sub-vector $\bm{\alpha^j}$ is treated as an SPC and decoded independently through a Wagner decoder. 
For every sub-vector, the index of the least reliable position is identified as $p^j = \argmin_{i}\{\alpha_i^j\}$, and the partial sum vector $\bm{\beta^j}$ is calculated through hard-decisions as
\begin{equation}
\beta_i^j =
  \begin{cases}
    \text{HD}(\alpha_i^j) \oplus \mbox{Parity} & \mbox{if } i = p_j \\
    \text{HD}(\alpha_i^j) & \mbox{otherwise}\mbox{,} 
  \end{cases} 
\end{equation}
where $\mbox{Parity}=\bigoplus \text{HD}(\alpha_i^j)$ for all $i$.
Finally, each element in the partial sum vector $\bm{\beta}$ of the G-PC node is calculated as 
\begin{equation}
\beta_i = \beta^{i \modulo N_p}_{\lfloor i/N_p \rfloor}~.
\end{equation}
It is worth noticing that the proposed decoding algorithm for G-PC nodes allows to reduce the decoding latency not only through SC tree pruning, but also by allowing decoder parallelization with a factor $N_p$, since the Wagner decoders are independent.

The even parity constraints imposed by the frozen bits in G-PC are the only independent ones present in the constituent code: thus, the proposed fast decoding technique is optimal. 
However, this technique can be applied also if other frozen bits are present, i.e. if some of the Rate-1 nodes are in fact Rate-C nodes with $\mbox{C} \simeq 1$. 
In this case, the even parity constraints are still valid, but other constraints brought by the inner frozen bits should be taken into account. 
We propose to ignore those additional constraints and apply Wagner decoding as if the node was a G-PC. 
In this case, we call this node a relaxed G-PC (RG-PC), and identify the frozen bits in the Rate-C node as additional frozen (AF) bits. 
The proposed decoding algorithm is suboptimal, and introduces a tradeoff between error-correction performance and decoding latency.

\subsection{Path Metric for List Decoding} \label{subsec:metric}

Fast list decoding of polar codes poses the question of how to compute the PM without descending the tree. It has been proven that fast decoding of Rate-0, Rate-1, Rep and SPC can be performed in list decoding as well, with the path metric computed exactly from the LLRs input to the node \cite{sarkis_list,hashemi_SSCL,hashemi_FSSCL,hashemi_SSCL_TCASI,hashemi_FSSCL_TSP}. 

In the same way, the proposed generalized fast decoding allows to compute exactly the path metrics at the top of the tree. Path metrics for G-Rep nodes are computed in two stages: the first one is relative to the Rate-C node, and is computed according to the decoding criterion of its particular frozen bit pattern. Once the Rate-C node has been decoded, the G-Rep node path metric calculation follows the same criterion of standard Rep nodes \cite{hashemi_SSCL_TCASI}:

\begin{equation*}
\PM_{\text{G-Rep}}=  \PM_{\text{Rate-C}} + \frac{1}{2} \sum_{i = 0}^{N_t/N_p-2}\Big(\sum_{j = 0}^{N_p-1} \sgn\left(\alpha_i^j\right)\alpha_i^j - \eta_{j}\alpha_i^j\Big) \text{,}
\end{equation*}
where $\bm{\eta} = 1-2\bm{\beta}$ is output by the Rate-C node, and $\bm{\alpha}$ is received from the parent of the G-Rep node.

Since G-PC nodes are compositions of parallel SPC nodes, the path metric at the top of the tree is computed in the same way \cite{hashemi_SSCL_TCASI}, but considering $N_p$ independent SPC nodes:
\begin{equation*}
\PM_{\text{G-PC}}=  \sum_{j = 0}^{N_p-1} \PM_{\text{SPC}_j} \text{.}
\end{equation*}
The same calculation applies to RG-PC nodes: while it is suboptimal, since it ignores the constraints imposed by the AF bits, it is still exact with respect to the same metric computed descending the tree and ignoring said additional constraints.

\section{Performance Analysis} \label{sec:perf}

In this section, we analyze the impact of generalized fast decoding on both SC and SCL in terms of both speed and error-correction performance.

\subsection{Speed}

\begin{table*}[t!]
\vspace{10pt}
\begin{center}

\caption{Generalized fast decoding time steps.}
\label{tab:GFDspeed}
\setlength{\extrarowheight}{1.7pt}
\begin{tabular}{c|c||c|c|c|ccc||c|c|c|ccc}
 
 \multirow{3}{*}{$N$} & \multirow{3}{*}{$R$} & \multicolumn{6}{c||}{SC} & \multicolumn{6}{c}{SCL}\\
 \cline{3-14}
  &  &  \multirow{2}{*}{Fast-SSC} & \multirow{2}{*}{+ G-Rep} & \multirow{2}{*}{+ G-PC} & \multicolumn{3}{c||}{+ RG-PC} &  \multirow{2}{*}{SSCL-SPC} & \multirow{2}{*}{+ G-Rep} & \multirow{2}{*}{+ G-PC} & \multicolumn{3}{c}{+ RG-PC}\\
  & &  &  &  & 1 AF & 2 AF & 3 AF &  &  &  & 1 AF & 2 AF & 3 AF \\ 
  \hline
  \hline
 \multirow{5}{*}{128} & $1/8$ & 31 & 28 & 28 & 26 & 22 & 17 & 51 & 47 & 47 & 42 & 34 & 33 \\
						 & $1/4$ & 61 & 60 & 54 & 54 & 42 & 42 & 98  & 96 & 96 & 78 & 66 & 66\\
					  & $1/2$ & 82 & 80 & 80 & 80 & 49 & 39 & 176 & 172 & 172 & 172 & 113 & 103\\
 					  & $2/3$ & 52 & 51 & 51 & 50 & 40 & 35 & 200 & 198 & 198  & 192 & 170 & 137  \\
  					  & $5/6$ & 55 & 54 & 42 & 34 & 25 & 20 & 247  & 245 & 175 & 142 & 129 & 124 \\
\hline
 \multirow{5}{*}{256} & $1/8$ & 116 & 114 & 114 & 104 & 96 & 78 & 127  & 125 & 124 & 114 & 106 & 96 \\
 					  & $1/4$ & 142 & 140 & 140 & 140 & 120 & 115 & 187  &  184 & 184 & 184 & 156 & 151\\
					 	 & $1/2$ & 113 & 111 & 108 & 107 & 85 & 75 & 323 & 317 & 312 &  307& 269 & 235 \\
 					  & $2/3$ & 115 & 114 & 105 & 100 & 75 & 57 & 408  & 402 & 370 & 355 & 318 & 285\\
  					  & $5/6$ & 79 & 75 & 72 & 72 & 64 & 45 & 476 & 468 & 455 & 455 & 440 & 358 \\
  					 \hline
 \multirow{5}{*}{512}  & $1/8$ & 116 & 109  & 109 & 107 & 92 & 82 &  194 & 188 & 182 & 176 & 156 & 134\\
					   & $1/4$ & 232 & 220 & 211 & 211 & 155 & 140 & 394  & 382  & 342 & 342 & 342 & 252 \\
					 & $1/2$ & 238 & 231 & 231 & 224 & 163 & 131 &  650  & 641 & 641 & 624 & 515 & 477\\
 					  & $2/3$ & 202 & 193 & 190 & 185 & 151 & 121 & 805  &  797 & 785 & 771 & 707 & 617\\
  					  & $5/6$ & 136 & 125 & 116 & 113 & 86 & 78 & 940  & 925 & 891 & 881 & 831 & 793 \\
  					 
\hline
 \multirow{5}{*}{1024}  & $1/8$ & 250 & 240 & 240 & 238 & 185 & 160 &  398  & 386  & 386 & 380 &309 & 276\\
   					  & $1/4$ & 353 & 344 & 344 & 344 & 269 & 224 & 712  &  702 & 697 & 697 & 589 & 496 \\
						 & $1/2$ & 420 & 405 & 405 & 401 & 311 & 256 & 1274 & 1251 & 1251 & 1241 & 1091 & 936 \\
 					  & $2/3$ & 344 & 335 & 334 & 334 & 254 & 211 & 1444  & 1432 & 1422 & 1397 & 1280 & 1174\\
  					  & $5/6$ & 232 & 224 & 215 & 202 & 173 & 141 & 1477  & 1470 & 1431 & 1350 & 1305 & 1195\\

\end{tabular}
\end{center}
\end{table*}

Table \ref{tab:GFDspeed} shows the number of time steps required to decode a set of polar codes with different lengths and rates, and different decoding algorithms. 
In particular, four code lengths are considered, namely $N=\{128,\,256,\,512,\,1024\}$, and five code rates $R=\{1/2,\,2/3,\,5/6,\,1/4,\,1/8\}$. All combinations of $N$ and $R$ have been constructed targeting the AWGN channel, and a noise standard deviation $\sigma=0.5$. Two baseline decoding algorithms are considered, i.e. Fast-SSC \cite{sarkis} and SSCL-SPC \cite{hashemi_SSCL_TCASI}; to each of these algorithms, the proposed generalized fast decoding is progressively applied, evaluating the impact on the decoding speed of G-Rep nodes first, then G-Rep and G-PC combined, and finally the three proposed node types, with increasingly high number of AF bits for RG-PC nodes.
The cost of decoding operations for the considered SC-based algorithms is computed as follows: both $f_t$ and $g_t$ operations have a cost of $1$ time step, regardless of the stage $t$. 
Rate-0 and Rate-1 nodes cost $1$ time step each, while Rep nodes and SPC nodes cost $2$ and $3$ time steps, respectively. 
G-Rep nodes have a cost of $1$ time step plus whatever is the cost of the Rate-C node, while both G-PC and RG-PC require $3$ time steps to be completed. 
These cost assumptions do not assume any kind of resource limitation. 
For SCL-based decoding, we instead assume a common structure in SCL decoders, in which the bit estimates memory structure implements a hardwired XOR tree so that partial sums relative to all SC tree stages are updated as soon as a bit is estimated  \cite{balatsoukas_SCL_HW,hashemi_SSCL_TCASI}. 
This implies that information bits need to be estimated one at a time. 
Moreover, every bit estimation is coupled with the path metric calculation and sorting, and selection of the surviving paths. 
Thus, given that the size of a node at stage $t$ is $2^t$, the cost of Rate-1 nodes rises to $2\times 2^t$, that of Rep nodes to $1+2^t$, and that of SPC nodes to $2\times 2^t-1$. 
In the same way, G-PC and RG-PC node require $1+2\times(\frac{2^t-1}{N_p})$ time steps. 
The cost of Rate-0 and G-Rep nodes is unchanged.

G-Rep nodes can be mostly found close to the imaginary border between the majority of unreliable bit-channels and the majority of reliable ones. 
Regardless of the code rate, their number is small, and the gain in terms of time steps limited. 
The size of G-Rep nodes tends to increase as the code length increases. 

As pointed out in Section \ref{subsec:GPC}, G-PC nodes revert to SPC nodes when $N_p=1$. 
Additional G-PC nodes where $N_p>1$ are not always found: this can be noticed by the fact that the number of time steps in the ``+ G-PC" columns in Table \ref{tab:GFDspeed} is sometimes unchanged from the ``+ G-Rep" columns. 
Nevertheless, the speedup brought by the fast decoding of G-PC nodes can be significant: for SC decoding, G-PC can save up to $21.8\%$ time steps with $N=128$, $R=5/6$, for a combined gain of $23.6\%$ with G-Rep nodes. 
With SCL decoding, the gain brought by G-PC nodes is larger, since $N_p$ SPC nodes of shorter length can be decoded in parallel: G-PC nodes save up to $28.3\%$ time steps, and up to $29.2\%$ when combined to G-Rep nodes.

Similar behavior can be observed for RG-PC nodes. 
With higher number of AF bits, the number of RG-PC nodes found in a code increases, and so does their time step saving. 
For SC decoding, RG-PC nodes with a single AF bit can save up to $14.5\%$ time steps with $N=128$, $R=5/6$, for a combined contribution with G-Rep and G-PC nodes of $38.2\%$. 
The combined gain can reach $54.5\%$ if the AF bits increase to $2$, and to $63.6\%$ with $3$ AF bits. 
The gain brought by RG-PC nodes in SCL decoding is larger in absolute value, but averagely smaller in percentage. 
With one AF bit, the time step gain can reach $18.4\%$, with $N=128$, $R=1/4$, for a combined contribution with G-Rep and G-PC nodes of $20.4\%$. 
The combined time step gain can instead reach $47.8\%$ and $49.8\%$ with 2 and 3 AF bits, respectively. 
As shown in the next Section, with a higher number of AF bits comes more significant error-correction performance loss.

The decoding speed can be further increased with respect to the results detailed in Table \ref{tab:GFDspeed} by ad-hoc code construction that maximizes the occurrence of the identified special nodes, as shown in \cite{GiardLowRate}.

\subsection{Error-Correction Performance}

\begin{figure}
  \centering
   \scalebox{0.98}{\begin{tikzpicture}
  \pgfplotsset{
    label style = {font=\fontsize{9pt}{7.2}\selectfont},
    tick label style = {font=\fontsize{7pt}{7.2}\selectfont}
  }

\begin{axis}[
	scale = 1,
    ymode=log,
    xlabel={$E_b/N_0$ [\text{dB}]}, xlabel style={yshift=0.4em},
    ylabel={BLER}, ylabel style={yshift=-0.75em},
    grid=both,
    ymajorgrids=true,
    xmajorgrids=true,
    grid style=dashed,
    mark options=solid,
    width=1\columnwidth, height=7cm,
    thick,
        xmin=0,
    mark size=3,
    legend style={
      anchor={center},
      cells={anchor=west},
      mark options=solid,
      column sep= 2mm,
      font=\fontsize{7pt}{7.2}\selectfont,
    },
    legend to name=ECP-1K,
    legend columns=2,
]

\addplot[
    color=black,
    dashed,
    mark=x,
    thick,
    mark size=3,
]
table {
0   1
0.5 0.996
1.0 0.94
1.5 0.741
2.0 0.406
2.5 0.131
3.0 0.0197785
3.5 0.00135252
4.0 2.21202e-005
};
\addlegendentry{Fast-SSC}

\addplot[
    color=black,
    mark=x,
    thick,
    mark size=3,
]
table {
0   1
0.5 0.982
1.0 0.967
1.5 0.499
2.0 0.171
2.5 0.035
3.0 0.00150096
3.5 3.79856e-005
};
\addlegendentry{SSCL-SPC}

\addplot[
    color=red,
    dashed,
    mark=triangle,
    thick,
    mark size=3,
]
table {
0.0 1
0.5 0.997
1.0 0.942
1.5 0.743
2.0 0.420
2.5 0.137
3.0 0.0200053
3.5 0.00140115
4.0 2.3601e-005
};
\addlegendentry{Fast-SSC + RG-PC, 1 AF}

\addplot[
    color=red,
    mark=triangle,
    thick,
    mark size=3,
]
table {
0   1
0.5 0.982
1.0 0.967
1.5 0.501
2.0 0.174
2.5 0.0361
3.0 0.0015768
3.5 3.83202e-005
};
\addlegendentry{SSCL-SPC + RG-PC, 1 AF}

\addplot[
    color=blue,
    dashed,
    mark=diamond,
    thick,
    mark size=3,
]
table {
0.0 1
0.5 0.997
1.0 0.95
1.5 0.777
2.0 0.486
2.5 0.208
3.0 0.0321232
3.5 0.0026101
4.0 6.2554e-005
};
\addlegendentry{Fast-SSC + RG-PC, 2 AF}

\addplot[
    color=blue,
    mark=diamond,
    thick,
    mark size=3,
]
table {
0   1
0.5 0.982
1.0 0.969
1.5 0.519
2.0 0.198
2.5 0.0455
3.0 0.0023021
3.5 7.10357e-005
};
\addlegendentry{SSCL-SPC + RG-PC, 2 AF}
\addplot[
    color=green,
    dashed,
    mark=o,
    thick,
    mark size=3,
]
table {
0.0 1
0.5 0.998
1.0 0.955
1.5 0.782
2.0 0.512
2.5 0.248
3.0 0.0460798
3.5 0.00474551
4.0 1.2022e-004
};
\addlegendentry{Fast-SSC + RG-PC, 3 AF}

\addplot[
    color=green,
    mark=o,
    thick,
    mark size=3,
]
table {
0   1
0.5 0.983
1.0 0.973
1.5 0.579
2.0 0.231
2.5 0.0516
3.0 0.0031297
3.5 1.087551e-004
};
\addlegendentry{SSCL-SPC + RG-PC, 3 AF}
\end{axis}
\end{tikzpicture}}
   \ref{ECP-1K}
  \\
  \vspace{2pt}
  \caption{BLER curves for $N=1024$, $R=1/2$. For SSCL-SPC, $L=4$, CRC length 16.}
  \label{fig:ECP-1K}
\end{figure}
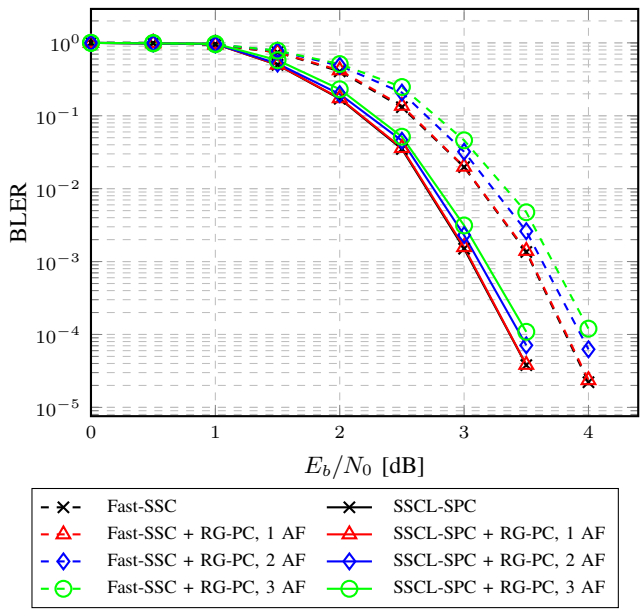

\begin{figure}
  \centering
   \scalebox{0.98}{\begin{tikzpicture}
  \pgfplotsset{
    label style = {font=\fontsize{9pt}{7.2}\selectfont},
    tick label style = {font=\fontsize{7pt}{7.2}\selectfont}
  }

\begin{axis}[
	scale = 1,
    ymode=log,
    xlabel={$E_b/N_0$ [\text{dB}]}, xlabel style={yshift=0.4em},
    ylabel={BLER}, ylabel style={yshift=-0.75em},
    grid=both,
    ymajorgrids=true,
    xmajorgrids=true,
    grid style=dashed,
    mark options=solid,
    width=1\columnwidth, height=7cm,
    thick,
    xmin=0,
    mark size=3,
    legend style={
      anchor={center},
      cells={anchor=west},
      mark options=solid,
      column sep= 2mm,
      font=\fontsize{7pt}{7.2}\selectfont,
    },
    legend to name=ECP-256,
    legend columns=2,
]

\addplot[
    color=black,
    dashed,
    mark=x,
    thick,
    mark size=3,
]
table {
0.0 0.841   
1.0 0.752
2.0 0.638
3.0 0.299
4.0 0.067
5.0 0.0112459
6.0 0.00137914
7.0 8.31117e-005
};
\addlegendentry{Fast-SSC}

\addplot[
    color=black,
    mark=x,
    thick,
    mark size=3,
]
table {
0.0 0.751   
1.0 0.414
2.0 0.202
3.0 0.042
3.5 0.0156687
4.0 0.00484136
4.5 0.00121963
5.0 0.000258041
5.5 2.70535e-005
};
\addlegendentry{SSCL-SPC}

\addplot[
    color=red,
    dashed,
    mark=triangle,
    thick,
    mark size=3,
]
table {
0.0 0.843   
1.0 0.761
2.0 0.648
3.0 0.328
4.0 0.104
5.0 0.0212459
6.0 0.00337914
7.0 2.8217e-004
};
\addlegendentry{Fast-SSC + RG-PC, 1 AF}

\addplot[
    color=red,
    mark=triangle,
    thick,
    mark size=3,
]
table {
0.0 0.754   
1.0 0.419
2.0 0.2242
3.0 0.053
3.5 0.021687
4.0 0.00674136
4.5 0.00181963
5.0 0.000408041
5.5 5.80535e-005
};
\addlegendentry{SSCL-SPC + RG-PC, 1 AF}

\addplot[
    color=blue,
    dashed,
    mark=diamond,
    thick,
    mark size=3,
]
table {
0.0 0.843   
1.0 0.763
2.0 0.651
3.0 0.369
4.0 0.154
5.0 0.032549
6.0 0.0052154
7.0 6.3777e-004
};
\addlegendentry{Fast-SSC + RG-PC, 2 AF}

\addplot[
    color=blue,
    mark=diamond,
    thick,
    mark size=3,
]
table {
0.0 0.771   
1.0 0.448
2.0 0.239
3.0 0.061
3.5 0.026910
4.0 0.00854136
4.5 0.00220963
5.0 0.000532041
5.5 9.23535e-005
};
\addlegendentry{SSCL-SPC + RG-PC, 2 AF}

\addplot[
    color=green,
    dashed,
    mark=o,
    thick,
    mark size=3,
]
table {
0.0 0.843   
1.0 0.768
2.0 0.671
3.0 0.391
4.0 0.174
5.0 0.039549
6.0 0.0077154
7.0 9.3777e-004
};
\addlegendentry{Fast-SSC + RG-PC, 3 AF}

\addplot[
    color=green,
    mark=o,
    thick,
    mark size=3,
]
table {
0.0 0.773   
1.0 0.454
2.0 0.242
3.0 0.065
3.5 0.030210
4.0 0.00949136
4.5 0.00250963
5.0 0.000602041
5.5 1.104535e-004
};
\addlegendentry{SSCL-SPC + RG-PC, 3 AF}
\end{axis}
\end{tikzpicture}}
   \ref{ECP-256}
  \\
  \vspace{2pt}
  \caption{BLER curves for $N=256$, $R=1/8$. For SSCL-SPC, $L=4$, CRC length 8.}
  \label{fig:ECP-256}
\end{figure}
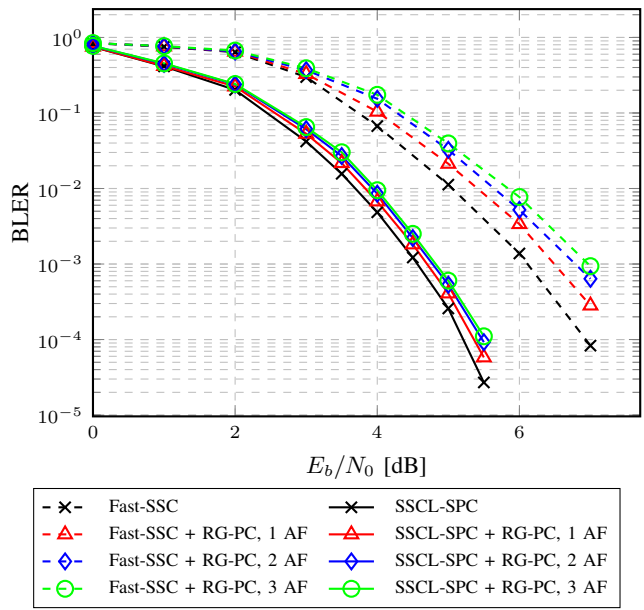

The proposed G-Rep and G-PC nodes do not impact the error-correction performance of the considered Fast-SSC and SSCL-SPC decoding algorithms. However, the approximation introduced by the RG-PC nodes can cause a performance loss. As an example, Fig. \ref{fig:ECP-1K} and Fig. \ref{fig:ECP-256} report the block error rate (BLER) curves for $N=1024$, $R=1/2$ and $N=256$, $R=1/8$, respectively. The SSCL-SPC curves have been obtained with a a list size $L=4$, and a CRC length of 16 for Fig. \ref{fig:ECP-1K} and of 8 for Fig. \ref{fig:ECP-1K}. It can be seen that as the number of AF bits increases, a larger error-correction performance degradation is observed. The entity of this degradation depends on the number of RG-PC nodes encountered, the length and rate of the code, and the effectiveness of the decoding algorithm. List-based decoding shows a higher degree of resilience to the RG-PC degradation in both cases, while the weaker code used in Fig. \ref{fig:ECP-256} suffers larger losses than its stronger counterpart in Fig. \ref{fig:ECP-1K}.

\section{Conclusion and Future Work} \label{sec:conc}

In this work, we introduced a generalized approach to fast decoding of polar codes. It identifies three multi-node subcode patterns that, along with including most existing subcodes, allow fast decoding of a wide variety of frozen and information bit patterns. Decoding rules are provided for any SC-based decoding algorithm, while fast path metric calculation for SCL is derived as well.
The proposed decoding approach is evaluated in terms of speedup and error correction performance against baseline fast decoding algorithms, over a wide set of code lengths and rates. Without any error-correction performance degradation, our technique shows up to $23.6\%$ and $29.2\%$ decoding latency gain with respect to fast SC and SCL decoding algorithms, respectively. These figures can rise up to $63.6\%$ and $49.8\%$ if a performance loss is accepted: the entity of the degradation depends on the combination of code and decoding algorithm parameters, and on the desired speedup.

The framework described in this work is not limited to polar codes; since the three identified subcodes are multi-node patterns, they valid for multi-kernel codes as well \cite{Gabry_MK}, that are constructed with combinations of kernels of different sizes. Future work foresees the evaluation of the effectiveness of the proposed generalized fast decoding to practical multi-kernel codes.

\bibliographystyle{IEEEtran}

\end{document}